\documentclass[12pt]{article}
\usepackage{amsmath}%\pdfoutput=1
\usepackage{cite}
\begin{document}
\begin{flushright}
KANAZAWA-17-01\\
June, 2017
\end{flushright}
\vspace*{1cm}
  
\renewcommand\thefootnote{\fnsymbol{footnote}}
\begin{center} 
{\Large\bf Symmetry restoration due to preheating\\
 and lepton number asymmetry}
\vspace*{1cm}

{\Large Daijiro Suematsu}\footnote[1]{e-mail:
~suematsu@hep.s.kanazawa-u.ac.jp}
\vspace*{0.5cm}\\

{\it Institute for Theoretical Physics, Kanazawa University, 
Kanazawa 920-1192, Japan}
\end{center}
\vspace*{1.5cm} 

\noindent
{\Large\bf Abstract}\\
We study a possible symmetry restoration due to the radiative effect 
of particles which are explosively produced in preheating after inflation.
As its application, we consider a scenario for leptogenesis 
based on the lepton number asymmetry generated in the right-handed 
neutrino sector through the inflaton decay. 
The scenario is examined in a one-loop radiative neutrino mass model
extended with singlet scalars.
             
\newpage
%%%%%%%%%%%%%%%%%%%%%%%%%%%%%%%%%%%
\setcounter{footnote}{0}
\renewcommand\thefootnote{\alph{footnote}}

\section{Introduction}
The existence of an inflationary expansion era in the early Universe 
seems to be justified from cosmic microwave background (CMB) 
observations \cite{uobs,planck15}. 
Inflation should be followed by some thermalization processes 
to realize an initial stage of the standard big-bang Universe.
Since inflation is usually considered to be caused by 
potential energy of 
inflaton which is a slow-rolling scalar field, this energy should 
be converted to radiation through certain reheating processes. 
Reheating is expected to be brought about by interactions of inflaton 
with contents of the standard model (SM) or others.
If we note that inflaton is usually identified with a singlet scalar,
we find that couplings with singlet fermions such as right-handed neutrinos 
could be one of their promising possibilities.
In that case, inflaton can also have quartic couplings with other scalar 
fields in general and the explosive particle production is expected 
to be induced through resonant instability called preheating \cite{preheat1}. 
Since preheating cannot convert the inflaton energy to radiation 
completely, inflaton should have a certain decay process to accomplish 
the reheating. From this point of view, the existence of the above mentioned 
coupling with the singlet fermions seems to be favored. 
Although the final reheating temperature is expected to be fixed through
this decay process, it has been suggested that preheating 
could play an important role in various phenomena which occurred at the 
early stage of evolution of the Universe, for example, 
symmetry restoration \cite{sym-rest1, sym-rest2}, 
baryogenesis \cite{pre-baryog}, phase transition \cite{pre-phase}, 
secondary inflation \cite{pre-sec-inf} and tachyonic preheating 
\cite{pre-tach}. We would like to propose a leptogenesis scenario
under the coexistence of such couplings.

In this paper, we discuss the symmetry restoration 
caused by the explosively produced particles through preheating 
\cite{sym-rest1, sym-rest2} as a basis of the supposing scenario. 
In particular, we focus on a possibility of the restoration of 
lepton number which is usually expected to be broken for the neutrino 
mass generation at low energy regions.
If the lepton number is restored due to such an effect 
at an early stage of the Universe, a new scenario of non-thermal 
leptogenesis might be considered as the origin of baryon number asymmetry 
in the Universe. As the concrete application of this scenario, 
we adopt an extended 
one-loop radiative neutrino mass model and examine such a possibility 
from a viewpoint of the connection with other phenomenology.

The remaining part of this paper is organized as follows.
In the next section, we briefly review the symmetry 
restoration caused by the radiative effect due to the explosively 
produced particles through preheating and then we apply it 
to two inflation scenarios.
In section 3, we discuss its application to the lepton number in a 
one-loop radiative neutrino mass model extended by singlet scalars.  
After the study of the neutrino mass generation and dark matter abundance 
in this model, we propose a scenario for non-thermal leptogenesis. 
We estimate an amount of lepton number asymmetry generated non-thermally 
through the inflaton decay by using the parameters which are consistent 
with neutrino oscillation data and dark matter abundance.
We summarize the paper in the final section. 

\section{Symmetry restoration via preheating}
\subsection{Preheating}
We briefly review the basics of the symmetry restoration due to 
preheating at first. 
The explosive particle production in the background of the inflaton 
oscillation is known as parametric resonance 
or preheating \cite{sym-rest1, sym-rest2}. 
Inflation is induced by a certain slow-roll potential
$V_{\rm inf}(\sigma)$ of a real scalar $\sigma$ called inflaton,
which is assumed to have a minimum at $\langle\sigma\rangle=0$.
If the inflaton $\sigma$ couples with 
a complex scalar $S$, the model could have a $U(1)$ symmetry.
We suppose that this $U(1)$ symmetry is spontaneously broken and then 
the potential is represented as\footnote{Since this potential looks like 
one of the hybrid inflation \cite{hybrid} with a waterfall field $S$, 
one might wonder if the scenario is based on the hybrid inflation. 
However, it should be noted that we assume that the potential during the 
inflation is dominated by the inflaton potential energy $V_{\rm inf}(\sigma)$ 
but not by the vacuum energy $\frac{\lambda_S}{4}u^4$.
The field value of the inflaton $\sigma$ is assumed to be of $O(M_{\rm pl})$
as the usual chaotic inflation. The amplitude of $\sigma$ is much larger
than that of $S$.} 
\begin{equation}
V(\sigma,S)= V_{\rm inf}(\sigma)
+\lambda_S\left(|S|^2-\frac{u^2}{2}\right)^2  + g_S\sigma^2|S|^2. 
\label{spot}
\end{equation}
The mass of $S$ could be expressed as $m_S^2=g_S\sigma^2-\lambda_S u^2$ 
during the slow-roll inflation. 
After the end of the inflation, the inflaton $\sigma$ 
oscillates around the potential minimum $\langle\sigma\rangle=0$. 
The oscillation is described by the equation 
\begin{equation}
\ddot\sigma+3H\dot\sigma+\frac{dV(\sigma)}{d\sigma}=0,
\label{eq1}
\end{equation}
where a dot stands for a time derivative and $S$ is assumed to stay initially 
at its local minimum $\langle S\rangle=\frac{u}{\sqrt 2}$. 
$H$ is the Hubble parameter given by
\begin{equation}
H^2=\frac{\frac{1}{2}\dot\sigma^2+V_{\rm inf}(\sigma)}{3M_{\rm pl}^2},
\end{equation}
where we use the reduced Planck mass 
$M_{\rm pl}=\frac{m_{\rm pl}}{\sqrt{8\pi}}$. 
The solution of eq.~(\ref{eq1}) might be represented by using the 
inflaton mass $\tilde m_\sigma$ as $\sigma(t)=\Sigma(t)\sin(\tilde m_\sigma t)$. 
Its amplitude $\Sigma(t)$ decreases due to the expansion 
of the Universe and rapidly approaches to its asymptotic value 
$\Sigma(t)=2\sqrt{\frac{2}{3}}\frac{M_{\rm pl}}{\tilde m_\sigma t}$.
At the first stage of this oscillation, 
the U(1) symmetry could be restored for a certain period
since the amplitude $\Sigma(t)$ is large enough 
to be $\Sigma^2(t)>\frac{\lambda_S}{g_S} u^2$.
Because of both the expansion of the Universe and the
production of $\sigma$ and $S$, which could happen depending 
on a self-coupling in $V_{\rm inf}(\sigma)$ and $g_S$,
the oscillation amplitude $\Sigma(t)$ decreases to result in 
$\Sigma^2(t)<\frac{\lambda_S}{g_S} u^2$.
The $U(1)$ symmetry seems to be broken at this period. 
However, the explosively produced $S$ could restore this symmetry.

In order to study the $S$ production under the background 
oscillation of $\sigma$, we introduce the shifted field $\tilde S$ 
around the symmetry broken vacuum $\langle S\rangle=\frac{u}{\sqrt 2}$.
It is expressed as 
$S=\langle S\rangle+\tilde S$ and $\tilde S=\frac{1}{\sqrt 2}(S_++iS_-)$.
The equation of motion for 
a quantum mode $S_{\pm p}$ with momentum $p(\equiv|{\bf p}|)$ is given as
\cite{preheat1}
\begin{equation}
\ddot S_{\pm p}+3H\dot S_{\pm p}+\omega_{\pm p}^2 S_{\pm p}=0.
\label{eq2}
\end{equation} 
The frequency $\omega_{\pm p}$ is defined as 
\begin{equation}
\omega_{\pm p}^2=\frac{p^2}{a^2}+g_S\sigma^2+ m_{S_\pm}^2,
\end{equation}
where $m_{S_+}^2=2\lambda_Su^2$ and $m_{S_-}^2=0$.\footnote{In the later study 
we introduce an additional mass term $m_S^2S^2$ for leptogenesis.
In that case, $m_{S_\pm}^2$ is replaced 
by the ones in eq.~(\ref{mscalar}). }
We note that $\omega^2_{\pm p}$ depends on $\sigma$ due to the last term 
of eq.~(\ref{spot}).
The scale parameter $a(t)$ satisfies the equation $\frac{\dot a}{a}=H$. 
The dynamics of $\sigma$ and the quantum scalar $S_\pm$ in the entire 
regime of interest can be treated by solving the coupled equations 
(\ref{eq1}) and (\ref{eq2}).
If the expansion of the Universe is neglected, 
eq.~(\ref{eq2}) is reduced to
\begin{equation}
S_{\pm p}^{\prime\prime}+(A_p-2q\cos 2z)S_{\pm p}=0,
\end{equation} 
where $A_p=\frac{p^2}{\tilde m_\sigma^2}+2q$, $q=\frac{g_S\Sigma^2}
{4\tilde m_\sigma^2}$ and $z=\tilde m_\sigma t$ is used. 
A prime represents the differentiation with respect to $z$.
This equation is known as the Mathieu equation whose solution is characterized 
by the stability/instability chart in the $(q, A_p)$ plane.  
The solution is expressed as 
$S_{\pm p}\propto \exp(\mu_{\pm p}^{(n)}z)$ 
by using a certain constant $\mu_{\pm p}^{(n)}$ which is fixed within the 
resonance bands of momenta $\Delta p^{(n)}$ labeled by an integer $n$. 
This could be interpreted to show the exponential growth of 
the number density of the produced particles such as 
$n_p(t)\propto \exp(2\mu_{\pm p}^{(n)}z)$ \cite{preheat1}.

However, if the effect of the expansion of the Universe is taken 
into account, the simple 
application of the stability/instability chart for the Mathieu equation 
is not allowed.
In that situation, the amplitude of the background field oscillation 
decreases and the momentum in the resonance bands cannot keep 
its position due to the red-shift effect. 
Thus, the existence of the parametric resonance in the expanding Universe
requires that the momenta in the resonance bands should not be 
red-shifted away from them before the sufficient particle 
production \cite{preheat1}.
The condition for its realization is known to be summarized 
as $q^2\tilde m_\sigma> H$.
If we note that $\left(\frac{H}{\tilde m_\sigma}\right)^{1/4}\simeq
\left(\frac{\Sigma}{M_{\rm pl}}\right)^{1/4}$ takes an almost stable
value of $O(1)$ for the first stage of the inflaton oscillation, 
this condition is found to be written as 
\begin{equation}
\sqrt{g_S}~\Sigma(t)>\sqrt{2}\tilde m_\sigma.
\label{cond1}
\end{equation}  
The occupation number $n_{\pm p}$ of this produced particle mode 
$S_{\pm p}$ can be estimated by using the solution $S_{\pm p}$ 
of eq.~(\ref{eq2}) as
\begin{equation}
n_{\pm p}=\frac{1}{2\omega_{\pm p}}\left(|\dot S_{\pm p}|^2
+\omega_{\pm p}^2|S_{\pm p}|^2\right)-\frac{1}{2}.
\label{ini-numb}
\end{equation}
For the estimation of this occupation number,
it is useful to use a typical momentum 
$p_\ast\simeq\sqrt{g_S^{\frac{1}{2}}\tilde m_\sigma \Sigma(t_0)}$ to find the 
resonance band. $\frac{p_\ast}{2}$ is expected to be contained 
in the resonance bands of $S_{\pm p}$ at time $t_0=\frac{\pi}{2\tilde m_\sigma}$ 
when $\sigma(t_0)=0$ is realized first after the inflaton starts 
the oscillation \cite{preheat1}. 

The above argument shows that the inflaton mass $\tilde m_\sigma$ is 
a crucial parameter in the preheating.
The potential $V_{\rm inf}(\sigma)$ is known to be constrained by the data
from the CMB observations. 
If we express the power spectrum of scalar perturbation 
as ${\cal P}_{\cal R}=A_S\left(\frac{k}{k_\ast}\right)^{n_s-1}$,
it suggests $A_S\simeq 2.43\times 10^{-9}$ at the time $t_\ast$ 
when the scale characterized by the wave number $k_\ast=0.002~{\rm Mpc}^{-1}$
exits the horizon \cite{uobs,planck15}. 
Since this condition can be rewritten by using a slow-roll parameter
$\epsilon$ which is defined by $\epsilon=\frac{M_{\rm pl}^2}{2}
\left(\frac{1}{V_{\rm inf}}\frac{dV_{\rm inf}}{d\sigma}\right)^2$ as 
\begin{equation}
\frac{V_{\rm inf}}{\epsilon}=(0.0275 M_{\rm pl})^4,
\label{cmb}
\end{equation} 
it gives a constraint on $V_{\rm inf}(\sigma)$.
For example, we may take the chaotic inflation 
$V_{\rm inf}(\sigma)=\frac{1}{2}\tilde m_\sigma^2\sigma^2$ although it is now
ruled out from the tensor-to-scalar ratio of the amplitude of 
the CMB power spectrum. 
Since $\epsilon\simeq \frac{1}{2N}$ is satisfied 
for the $e$-foldings $N$ in this example, it imposes 
$\tilde m_\sigma\simeq 1.5\times 10^{13}$~GeV for $N_\ast=60$ which stands 
for the $e$-foldings from $t_\ast$ to the end of inflation.  
However, if $V_{\rm inf}(\sigma)$ is described by different functions 
at the inflation era, the inflaton mass $\tilde m_\sigma$ might not be 
constrained in the same way.
We will focus our study on such examples.

These particles $S_\pm$ produced through the preheating are known to induce
the symmetry restoration \cite{sym-rest1, sym-rest2}.
In order to describe it, we consider quantum corrections 
brought about by the produced $S_\pm$ to the potential of $S$. 
During the inflaton oscillation, the effective potential for $S$ 
may be represented as
\begin{equation}
V_{\rm eff}(S)=\lambda_S\left(|S|^2-\frac{u^2}{2}\right)^2
+g_S\sigma^2|S|^2+V_1^0(S)
+V_1^f(S).
\label{effectp}
\end{equation}
$V_1^0(S)$ is the ordinary zero-temperature one-loop potential and 
$V_1^f(S)$ comes from the one-loop contribution caused by the particles
$S_\pm$ produced explosively through the preheating. 
Their momentum distribution is assumed to be described by a function $f(p)$.
Here we use the formalism given in \cite{therm} to estimate $V_1^f(S)$.
This is because the distribution function $f(p)$ is not the one in the thermal 
equilibrium and then the usual imaginary time formalism cannot be used.
The free propagator of $S_\pm$ in this formalism is written as 
a $2\times 2$ matrix and the one-loop effective potential $V_1$ can be 
given by using its (11)-component $D_{11}$. 
Following the procedure given in Appendix B of \cite{dj}, $V_1$ satisfies
\begin{equation}
\frac{dV_1}{d\bar m_{S_\pm}^2}=\frac{1}{2}\int \frac{d^4\bar p}{(2\pi)^4}
D_{11}(\bar p), 
\quad
D_{11}(\bar p)=\frac{i}{\bar p^2-\bar m_{S_\pm}^2+i\varepsilon}
+2\pi f(p)\delta(\bar p^2-\bar m_{S_\pm}^2),
\label{veff}
\end{equation}
where $\bar m_{S_\pm}$ is the field dependent mass of $S_\pm$. It is 
expressed as $\bar m_{S_\pm}^2=2\lambda_S\left(|S|^2\pm\frac{u^2}{2}\right)$.
$V_1^0(S)$ and $V_1^f(S)$ in eq.~(\ref{effectp}) come from the first and 
the second term in $D_{11}(\bar p)$, respectively.
For simplicity, we assume that the momentum distribution of 
the produced $S_\pm$ is written as $f(p,t)=A(t)p\delta(p-p_m)$ 
taking account of the red-shift effect.
In the expanding Universe, we find $V_1^f(S)$ by solving 
eq.~(\ref{veff}) as \cite{sym-rest2} 
\begin{equation}
V_1^f(S)=\int \frac{d^3p}{(2\pi a)^3}f(p)\left(\sqrt{p^2+\bar m_{S_+}^2}
+\sqrt{p^2+\bar m_{S_-}^2}\right)
\simeq \frac{A(t)\lambda_S p_m^2}{\pi^2a^3}|S|^2,
\end{equation}
where we use $p_m\gg\bar m_{S_\pm}$ and omit both $|S|$ independent terms and
higher order terms than $|S|^2$ in the last equality. 
The physical number density $n_{S}(t)$ obtained from the 
present distribution function $f(p,t)$ is expressed by taking account of the
Universe expansion as\footnote{Following 
the detailed analysis of the preheating in \cite{preheat1},
$A(t)$ in the distribution function $f(p)$ might be approximated 
by using $\mu$ which characterizes the particle production rate fixed 
by the model parameters. In that case, the number density of $S$ 
at time $t$ could be estimated as 
$n_S(t)\sim \frac{p_m^3}{4\pi^2a(t)^3}e^{2\mu\tilde m_\sigma t}$.} 
\begin{equation}
n_{S}(t)=\frac{p_m^3A(t)}{2\pi^2 a(t)^3}.
\label{number}
\end{equation}

The preheating is expected to end at $t\simeq t_f$ when 
condition (\ref{cond1}) is violated. It might be roughly estimated as 
$t_f\simeq\ 1.6\frac{\sqrt{g_S}M_{\rm pl}}{\tilde m_\sigma^2}$.
Since no explosive production of $S_\pm$ is expected after $t_f$,
the maximum number density is determined
as $n_{S}(t_f)$ by using eq.~(\ref{number}). 
The produced quanta are monotonically diluted by the expansion 
of the Universe after that.
If we take account of it and use the above effective potential whose 
dominant one-loop contribution comes from $V_1^f(S)$,
we find that the effective mass $\tilde m_{S}^2$ of $S$ at $t>t_f$ could be 
expressed as 
\begin{equation}
\tilde m_{S}^2(t)= g_S\langle\sigma\rangle^2+\lambda_S\left(- u^2 
+ \frac{2n_{S}(t_f)}{p_ma(t)^3}\right).
\label{effectivemass}
\end{equation}
Even when the amplitude of the inflaton $\sigma$ becomes small,
the last term induced by the quantum effect of $\tilde S$ could 
make $\tilde m_S^2$ positive and then the $U(1)$ symmetry is restored
in such a case.
This symmetry restoration could be kept until the time $t$
as long as the condition 
\begin{equation}
\frac{n_{S}(t_f)}{a(t)^3}>p_m \frac{u^2}{2} 
\label{smass}
\end{equation}
is satisfied.
If we impose it until the time when 
the reheating completes, the condition (\ref{smass}) could be rewritten as
\begin{equation}
 n_{S}(t_f)>p_m\frac{u^2}{2}\left(\frac{t_R}{t_f}\right)^2=\frac{2}{9}p_m u^2
\left(\frac{1}{\Gamma_\sigma t_f}\right)^2,
\label{cond2}
\end{equation}
where the matter dominated expansion $H=\frac{2}{3}t^{-1}$ 
is assumed from $t_f$ to the end of reheating. 
The reheating completion time $t_R$ could be fixed from $H\simeq \Gamma_\sigma$
by using the inflaton decay width $\Gamma_\sigma$.

In the next part, we numerically estimate the occupation number 
of the produced particles in two inflation models. 
It can be proceeded by solving numerically the above coupled equations 
(\ref{eq1}) and (\ref{eq2}) for $\sigma$, $S_{\pm p}$, and the Hubble 
equation $\frac{\dot a}{a}=H$ for suitable initial values.
In the equation of motion of $\sigma$, they could be fixed 
at $\sigma=\sigma_c$ and also 
$\dot\sigma|_{\sigma=\sigma_c}\simeq 0.8\sqrt{V_{\rm inf}}$, where
$\sigma_c$ is taken as an inflaton value at the end of inflation.
The latter could be derived by using the slow-roll equation 
$3H\dot\sigma\simeq -\frac{dV_{\rm inf}}{d\sigma}$.
On the other hand, if we use eq.~(\ref{ini-numb}), we could find the
appropriate initial condition for eq.~(\ref{eq2}).
At the initial stage of $\sigma$ oscillation, $n_{\pm p}=0$ should be 
satisfied. From this, we adopt $S_{\pm p}=\frac{1}{\sqrt{\omega_{\pm p}}}$ 
for $\dot S_{\pm p}=0$ as the initial condition.

\subsection{Two inflation scenarios} 
We study the symmetry restoration due to this particle production 
in two concrete inflation models here. 
We consider that the slow-roll inflation is caused by the inflaton potential
$V_{\rm inf}(\sigma)$ which is expressed as
\begin{equation}
V_{\rm inf}(\sigma)=\left\{\begin{array}{ll}   
V_{\rm I}(\sigma) & \quad \sigma>\sigma_c,  \\
\frac{1}{2}\tilde m_\sigma^2\sigma^2  & \quad  \sigma<\sigma_c. \\
\end{array}\right.
\label{infpot}
\end{equation}
If we tune the model parameters appropriately, 
the inflaton potential is expected to transit in this way
between inflation time and post inflation time. 
The slow-roll inflation is considered to be induced by $V_I$ and 
end at $\sigma\simeq\sigma_c$ where the slow-roll condition is violated 
to be $\epsilon\simeq 1$. 
At the post inflation era, the potential is supposed to be approximated as
$\frac{1}{2}\tilde m_\sigma^2\sigma^2$ before the reheating.
For example, if the $\kappa$ term dominates the potential for large $\sigma$
in $V_{I}=\frac{\kappa}{4}\sigma^4+\frac{1}{2}\tilde m_\sigma^2\sigma^2$, 
the CMB condition (\ref{cmb}) constrains $\kappa$ but not 
the value of $\tilde m_\sigma$ directly.  
Since $\sigma$ reduces its value as a result of the expansion, both terms 
can become equal soon at a certain time $t_e$ much before the reheating 
time $t_R$. If we take into account that both $t_e$ and $t_R$ are roughly 
estimated as $t_e\sim \sqrt{\frac{2\kappa}{3}}\frac{M_{\rm pl}}{\tilde m_\sigma}$
and $t_R\sim 2g_\ast^{-1/2}\frac{M_{\rm pl}}{T_R^2}$,
we find that $t_e\ll t_R$ could be possible for 
$\frac{T_R}{\tilde m_\sigma}\ll 0.5\kappa^{-1/4}$.    
In the following part, we consider two examples for this kind of 
inflaton potential, which
can satisfy the present data of the CMB tensor-to-scalar ratio \cite{planck15}. 

\noindent
{\it Model (a)} 

We consider a real scalar $\sigma$ whose Jordan frame potential is
written as $V(\sigma)=\frac{\kappa}{4}\sigma^4+
\frac{1}{2}\tilde m_\sigma^2\sigma^2$ and the $\kappa$ term is assumed to be
dominant for large values of $\sigma$. It is also assumed to have 
a non-minimal coupling  $\frac{\xi}{2}\sigma^2 R$ 
with Ricci scalar \cite{nonm-inf,higgsinf,sinfl,bks1}. 
In the Einstein frame, 
a canonically normalized field $\chi$ can be defined by
\begin{equation}
\frac{d\chi}{d\sigma}= 
\frac{\left[1+\left(\xi+6\xi^2\right)\frac{\sigma^2}{M_{\rm pl}^2}\right]^{1/2}}
{1+\frac{\xi\sigma^2}{M_{\rm pl}^2}},
\label{ein}
\end{equation}
and the scalar potential can be written as
\begin{equation}
V_{\rm inf}=\frac{1}{\Omega^4}\left(\frac{\kappa}{4}\sigma^4
+\frac{1}{2}\tilde m_\sigma^2\sigma^2\right), \qquad
\Omega^2=1+\frac{\xi\sigma^2}{M_{\rm pl}^2}.
\end{equation}
By using eq.~(\ref{ein}), $\sigma$ is found to be related to $\chi$ as 
$\sigma\propto\exp\left(\frac{\chi}{\sqrt{6+\frac{1}{\xi}}M_{\rm pl}}\right)$ 
at $\sigma\gg\frac{M_{\rm pl}}{\sqrt{\xi}}$ and $\chi$ reduces to $\sigma$
at $\sigma\ll\frac{M_{\rm pl}}{\sqrt{\xi}}$.
Thus, if we assume that 
$\frac{\kappa}{4}\sigma^4>\frac{1}{2}\tilde m_\sigma^2\sigma^2$ is satisfied at 
$\sigma_c=\frac{M_{\rm pl}}{\sqrt{\xi}}$,
$V_{\rm I}$ is found to be represented as   
$V_{\rm I}(\sigma)\simeq \frac{\kappa M_{\rm pl}^4}{4\xi^2}$ at 
$\sigma> \sigma_c$ and eq.~(\ref{infpot}) is also realized 
at $\sigma< \sigma_c$.
In this model, the slow-roll parameters $\epsilon$ and the $e$-foldings $N$ are
expressed approximately as $\epsilon=\frac{4M_{\rm pl}^4}{3\xi^2\sigma^4}$ and 
$N=\frac{3\xi\sigma^2}{4 M_{\rm pl}^2}$.
Thus, the initial conditions for the $\sigma$ oscillation are found to 
be summarized as
\begin{equation}
\sigma_c\simeq\left(\frac{4}{3}\right)^{1/4}\frac{M_{\rm pl}}{\sqrt\xi}, \qquad
\dot\sigma|_{\sigma=\sigma_c}\simeq0.4\frac{\sqrt\kappa}{\xi}M_{\rm pl}^2,
\end{equation}
if we use $\epsilon\simeq 1$ at $\sigma=\sigma_c$ and 
$\dot\sigma|_{\sigma=\sigma_c}\simeq 0.8\sqrt{V_{\rm I}}$.

From the CMB constraint (\ref{cmb}),
we find 
\begin{equation}
\kappa \simeq 1.7\times 10^{-6} \frac{\xi^2}{N_\ast^2}.
\end{equation}
Since these parameters have no phenomenological constraint differently 
from the Higgs inflation, we can take a value of $\xi$ freely in this study. 
We note that there appears no unitarity problem related to 
the inflation even in that case. Here we use $\xi=10$ as a moderate value.
The parameters relevant to the estimation of the particle production
are fixed for this value of $\xi$. 

\begin{figure}[t]
\begin{center}
\begin{tabular}{ccccccc}\hline
 & $g_S$ &$\kappa$  &$\tilde m_\sigma~({\rm GeV})$ & $\sigma_c~({\rm GeV})$ 
& $p_m~({\rm GeV})$ & $n_{\pm p}^c(t_f)$ \\ \hline
Model (a)&  $1.8\times 10^{-8}$&$4.7\times 10^{-8}$ & $3\times 10^{12}$ &
$8.3\times 10^{17}$ & $9.89\times 10^{13}$ &  $1.4\times 10^{8}$   \\ 
Model (b)&$4.3\times 10^{-8}$& $1.2\times 10^{-7}$ & $4\times 10^{12}$ &
$7.2\times 10^{17}$ & $1.26\times 10^{14}$ &  $4.2\times 10^{8}$  
\\ \hline
\end{tabular}
\end{center}
\vspace*{5mm}

{\footnotesize {\bf Table~1}~~Parameters used in the numerical study.
These correspond to $q=350$ and $\Sigma\simeq \sigma_c$ at the end 
of inflation. $p_m$ is the momentum at which the number density $n_{\pm p}$ 
takes a maximum. $N_\ast=60$ is assumed here. If we use 
$t_f\simeq 1.6\frac{\sqrt{g_S}M_{\rm pl}}{\tilde m_\sigma^2}$, $t_f$ 
is estimated as 28 in Model (a) and 32 in Model (b) in a 
$\frac{2\pi}{\tilde m_\sigma}$ unit. }
\end{figure}

\noindent
{\it Model (b)} 

We consider a complex scalar $\sigma$ whose potential is 
expressed as  \cite{bks2}
\begin{eqnarray}
V_{\rm inf}&=&V_I+ \tilde m_\sigma^2\sigma^\dagger\sigma+
\frac{1}{2}m_\sigma^2\sigma^2
+\frac{1}{2}m_\sigma^2\sigma^{\dagger 2}, \nonumber \\
V_{\rm I}&=&\kappa(\sigma^\dagger \sigma)^2\left[1+ 
\alpha\left\{\left(\frac{\sigma}{M_{\rm pl}}\right)^2
\exp\left(i\frac{\sigma^\dagger \sigma}{\Lambda^2}\right)+
\left(\frac{\sigma^\dagger}{M_{\rm pl}}\right)^2
\exp\left(-i\frac{\sigma^\dagger \sigma}{\Lambda^2}\right)\right\}\right] 
\nonumber\\
&=&\frac{\kappa}{4}\varphi^4
\left[1 + 2\alpha\left(\frac{\varphi}{\sqrt 2M_{\rm pl}}\right)^2
\cos\left(\frac{\varphi^2}{2\Lambda^2}+2\theta\right)\right] \nonumber \\
&\equiv& \frac{\kappa}{4}\varphi^4+ 
V_b\cos\left(\frac{\varphi^2}{2\Lambda^2}+2\theta\right),
\label{bpot} 
\end{eqnarray}
where $\sigma=\frac{1}{\sqrt 2}\varphi e^{i\theta}$.
If $V_b\cos\left(\frac{\varphi^2}{2\Lambda^2}\right)~{^<_\sim}~ 
\frac{\kappa}{4}\varphi^4$ is satisfied at $\varphi\simeq\varphi_c$, 
$V_{\rm inf}$ could be approximated by $V_I$ at $\varphi>\varphi_c$.
In that case, the inflation is induced through the flat direction of 
$V_{\rm I}$ which is represented by the inflaton $\chi$ constrained along the 
polar angle direction as long as $\sigma$ stays at the local minimum in
the radial direction \cite{mc}. The inflaton $\chi$ is defined as 
\begin{equation}
d\chi=\left[ 1 + \frac{1}{\varphi^2}\left(\frac{d\varphi}{d\theta}\right)^2
\right]^{1/2}\varphi d\theta
=\left[1+4\left(\frac{\Lambda}{\varphi}\right)^4\right]^{1/2}\varphi d\theta,
\label{inflaton}
\end{equation}
where in the second equality we use a fact that this constrained 
path satisfies $\frac{\varphi^2}{2\Lambda^2}+2\theta=(2m+1)\pi$ for 
an integer $m$.
Since the $e$-foldings $N$ and the slow-roll parameter $\epsilon$ 
can be approximately estimated as\footnote{The contribution 
from $V_b$ to these values is omitted in these approximated expressions 
since it is sub-dominant. Even if we use these formulas, 
the results are not affected in the present study.}
\begin{equation}
N\simeq\frac{1}{12}\left(\frac{\varphi}{\sqrt 2M_{\rm pl}}\right)^6
\left(\frac{M_{\rm pl}}{\Lambda}\right)^4, \qquad
\epsilon\simeq 4\left(\frac{\sqrt 2M_{\rm pl}}{\varphi}\right)^6
\left(\frac{\Lambda}{M_{\rm pl}}\right)^4,
\label{epsilon}
\end{equation}
the $e$-foldings $N$ and the slow-roll parameter $\epsilon$ are related 
to each other as $\epsilon\simeq\frac{1}{3N}$.

In this model, the single field slow-roll inflation picture cannot be applied
at the final stage of inflation since the inflaton $\chi$ defined by
eq.~(\ref{inflaton}) could not describe well the motion of $\sigma$
which deviates from the local minimum in the radial direction.
However, both $3H\dot \chi\simeq -\frac{dV_{\rm I}}{d\chi}$ and 
$\frac{1}{2}\dot \chi^2\simeq V_b$ are considered to be satisfied 
simultaneously at the end of inflation.  
If we use these conditions approximately, we can estimate the value 
of $\varphi$ at the end of inflation as 
\begin{equation}
\frac{\varphi_c}{\sqrt 2 M_{\rm pl}}\simeq 
\left(\frac{2}{3\alpha}\right)^{1/8}
\left(\frac{\Lambda}{M_{\rm pl}}\right)^{1/2}.   
\end{equation}
Since the inflaton could go over the potential barrier $V_b$ at 
$\varphi\simeq \varphi_c$, 
the complex scalar $\sigma$ cannot be kept in the constrained path 
and the components $\sigma_{1,2}$ of 
$\sigma(\equiv\frac{1}{\sqrt 2}(\sigma_1+i\sigma_2))$ could be considered
to oscillate in the approximated potential,
\begin{equation} 
V_{\rm inf}\simeq\frac{1}{2}
m_{+\sigma}^2\sigma_1^2+\frac{1}{2}m_{-\sigma}^2\sigma_2^2, \qquad
m_{\pm\sigma}^2=\tilde m_\sigma^2\pm m_\sigma^2
\end{equation}
at the region $\varphi<\varphi_c$. 
Since the potential for $\sigma_1$ and $\sigma_2$ is not
the same due to the existence of $m_\sigma^2$, the supposed $U(1)$ symmetry
could be violated in this part.\footnote{We come back to this point later 
to study the generation of the lepton number asymmetry through 
the inflaton decay. We assume $m_\sigma=0.3\tilde m_\sigma$ 
in the numerical study.} 
We note that the coupling $g_S\sigma^\dagger\sigma S^\dagger S$ relevant to 
the particle production is
written as $\frac{g_S}{2}(\sigma_1^2+\sigma_2^2)S^\dagger S$. 
The initial conditions for the oscillation of $\sigma_{1,2}$
at $\varphi=\varphi_c$ are found to be expressed as
\begin{equation}
\sigma_1= \varphi_c, \qquad \sigma_2=0, \qquad \dot \sigma_1=0, \qquad 
\dot \sigma_2=2\sqrt\alpha~\frac{\tilde m_\sigma\Lambda^2}{M_{\rm pl}}. 
\end{equation}
If we impose the CMB constraint (\ref{cmb}) on this model,
we find that $\kappa$ should satisfy the condition
\begin{equation}
\kappa\simeq 3.6\times 10^{-8}\frac{1}{N_\ast^{5/3}}
\left(\frac{M_{\rm pl}}{\Lambda}\right)^{8/3}.
\end{equation}
Parameters in the potential (\ref{bpot}) are adopted as
$\alpha=1.1$ and $\frac{\Lambda}{M_{\rm pl}}=0.05$ for $N_\ast=60$, 
which can explain the tensor-to-scalar ratio of the CMB perturbation
presented by Planck \cite{bks2}.

\input epsf
\begin{figure}[t]
\begin{center}
\epsfxsize=7.5cm
\leavevmode
\epsfbox{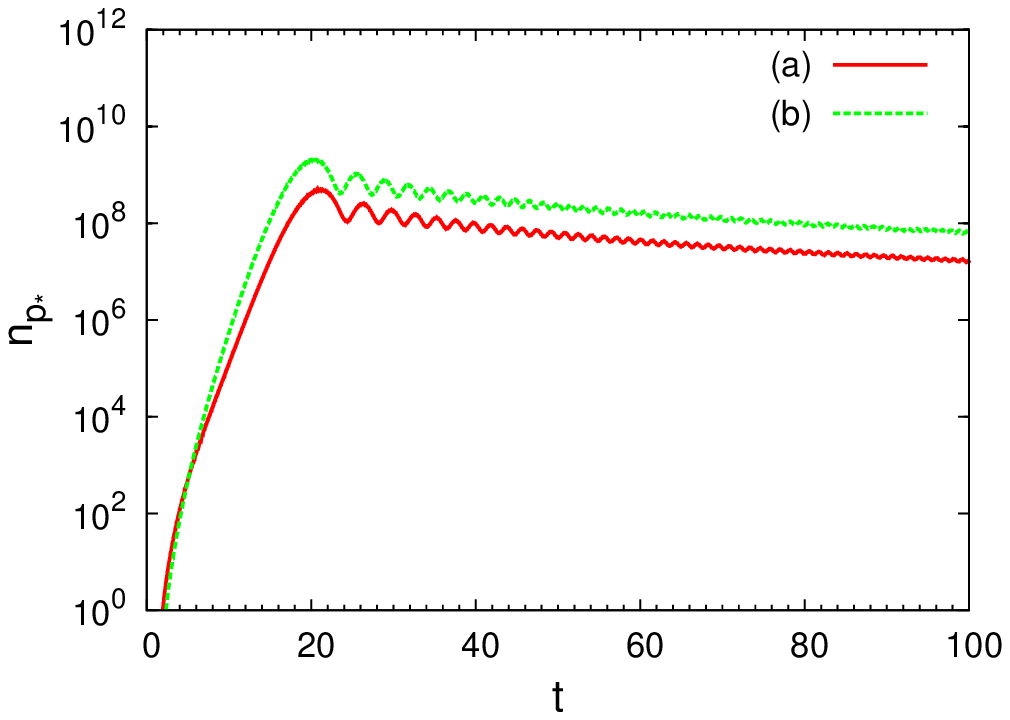}
\hspace*{5mm}
\epsfxsize=7.5cm
\leavevmode
\epsfbox{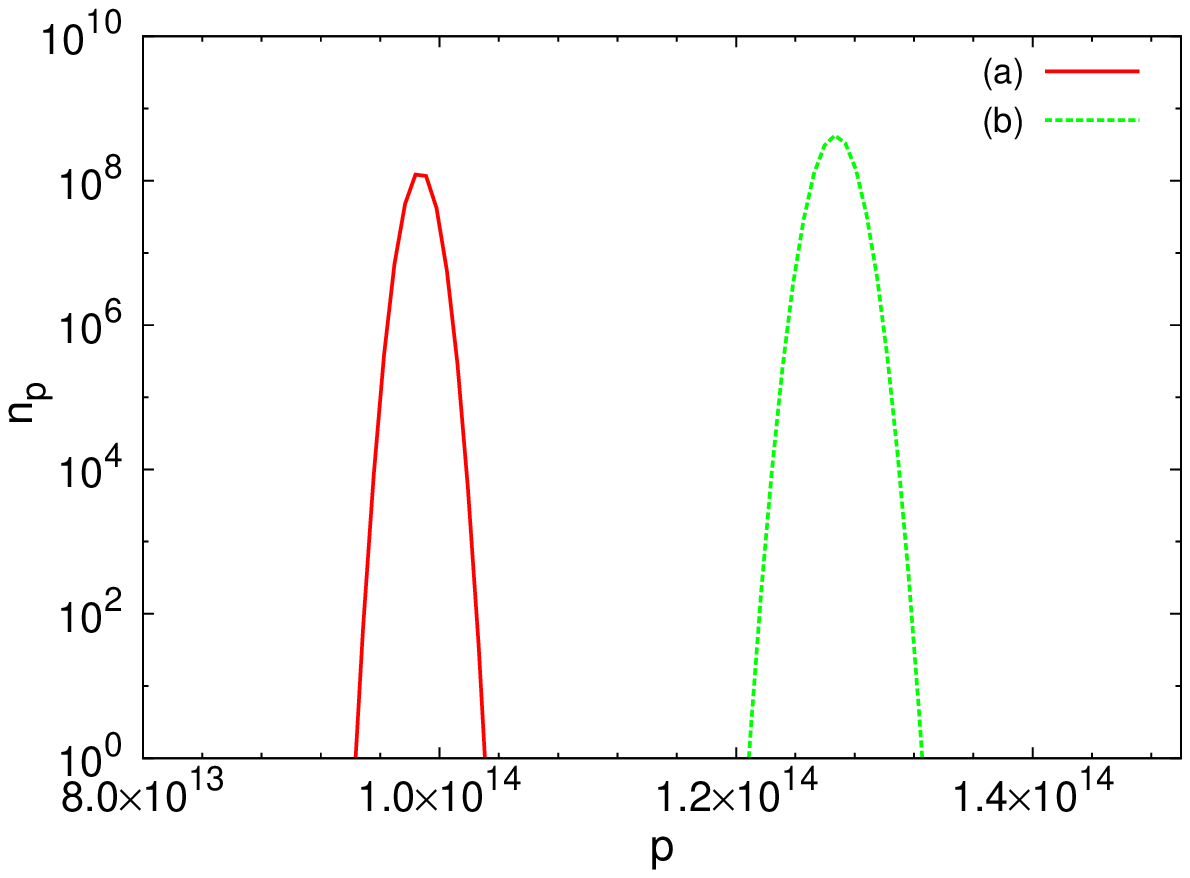}
\end{center}
%\vspace*{-3mm}

{\footnotesize {\bf Fig.~1}~~ Left: The evolution of the physical 
number density of a momentum mode $S_{\pm p_\ast}$ with a typical momentum 
$p_\ast$ which characterizes the position of a resonance band. 
It can be fixed at $p_\ast=\sqrt{g_S^{1/2}\tilde m_\sigma\Sigma(t_0)}$.
The time $t$ is taken as a $\frac{2\pi}{\tilde m_\sigma}$ unit.
Right: The momentum distribution $n_p$ of the produced particle $S_\pm$.
These curves are fixed by the Gaussian fitting to the numerical data points.
A unit of the momentum $p$ is taken to be GeV.
In both panels, labels (a) and (b) stand for the models discussed 
in the text and shown in Table~1. }
\end{figure}

Now we present results of the numerical study for the resonant 
$S_\pm$ production in the framework defined by eq.~(\ref{spot}), 
in which $V_{\rm inf}$ is taken as the above ones.
Parameters used in this study are listed in Table~1 for each model. 
In the left panel of Fig.~1, the number density $n_{\pm p_\ast}$ of the
momentum mode $S_{\pm p_\ast}$ generated in the preheating is shown for
each model. This figure shows that the exponential particle production 
continues from the end of inflation to the time $t_f$. 
As discussed in the previous part, the particle production stops there
since the condition (\ref{cond1}) is violated due to the 
red shift of the momentum and the decrease of the inflaton amplitude. 
After the end of preheating $t_f$, the number density $n_{\pm p_\ast}$ 
decreases monotonically due to the expansion of the Universe.
In the right panel of Fig.~1, the distribution of the produced 
momentum mode is plotted. It is obtained by applying the Gaussian fit 
to the numerical data points for several values of momentum $p$. 
The number density of $S_\pm$ obtained from the integration of 
this fitting function is found to be nicely approximated 
by $n_{\pm S}=\frac{p_m^3}{64\pi^2}n_{\pm p_m}$, where $n_{\pm p_m}$ stands for a
peak value of $n_{\pm p}$ realized at $p=p_m$. 
This suggests that we can put $A(t_f)=\frac{1}{32}n_{\pm p_m}(t_f)$ in
the previously assumed distribution function.
We use these results in the analysis of the symmetry restoration
in the model defined by eq.~(\ref{spot}). 
In the following study, the parameters contained in the potential 
are fixed as
\begin{equation}
u= 1.4\times 10^{15}~{\rm GeV}, \qquad \lambda_S=2.5\times 10^{-11}. 
\label{vev}
\end{equation}
The condition (\ref{smass}) for the symmetry restoration is found to 
be easily satisfied at $t_f$ when the preheating ends. 
It is crucial for the study of the related physics to know how long 
this symmetry restoration is kept. 

\begin{figure}[t]
\begin{center}
\epsfxsize=7.5cm
\leavevmode
\epsfbox{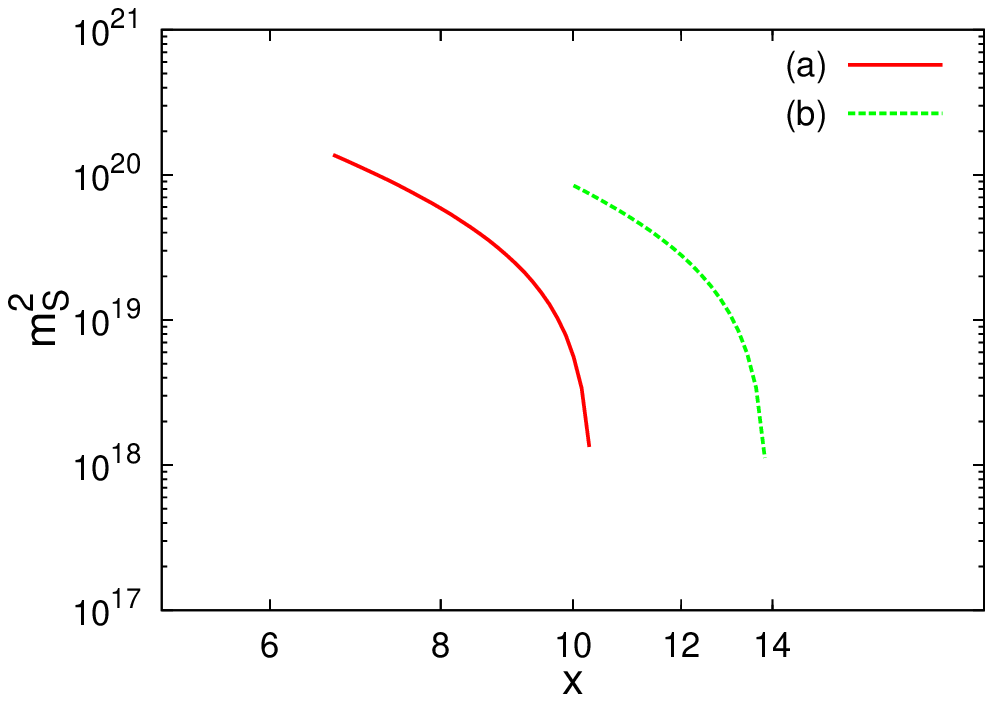}
\hspace*{5mm}
\epsfxsize=7.5cm
\leavevmode
\epsfbox{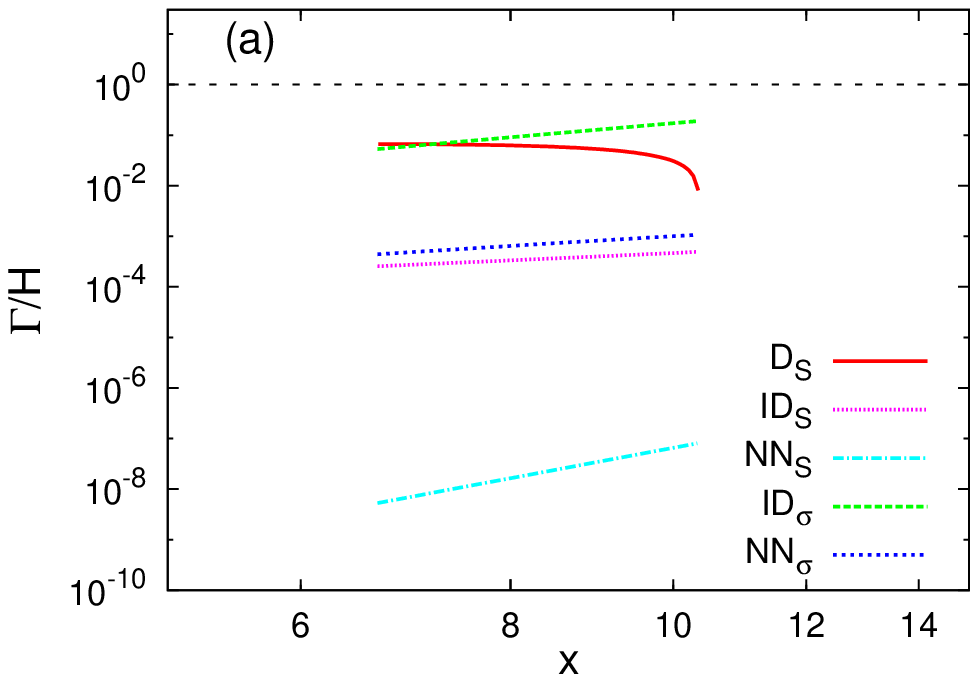}
\end{center}
%\vspace*{-3mm}

{\footnotesize {\bf Fig.~2}~~ 
Left: The effective squared mass $\tilde m_S^2$ in a GeV unit as a function 
of $x\left(\equiv\frac{M_1}{T}\right)$. The right-handed neutrino mass 
is fixed at $M_1=10^{14}$~GeV. 
The labels (a) and (b) stand for the model shown in Table~1.
Right: The reaction rates $\frac{\Gamma}{H}$ for the lepton number 
violating processes in Model (a). 
The $S$ decay, the inverse decay of $S$ and $\sigma$, 
and the $NN$ scattering mediated by $S$ and $\sigma$ are labeled 
by $D_S$, ID$_{S,\sigma}$, and NN$_{S,\sigma}$, respectively.   
The left end point of each line corresponds to the reheating 
temperature $T_R$.}
\end{figure}

We discuss this problem from a viewpoint to make this symmetry 
restoration applicable for a new type of non-thermal leptogenesis 
scenario. For that purpose, we introduce right-handed neutrinos to
fix the reheating process and impose them to couple 
with both $\sigma$ and $S$ through
\begin{equation}
-{\cal L}= \zeta_i\sigma \bar N_i^cN_i + \zeta_i^\ast\sigma \bar N_iN_i^c
+y_iS\bar N_i^cN_i + y_i^{\ast}S^\dagger\bar N_iN_i^c.
\label{fmodel}
\end{equation}
If we assign the lepton number $N_i$ and $S$ such as
$L(N_i)=1$ and $L(S)=-2$, the U(1) symmetry discussed above
could be identified with this lepton number.
The reheating is finally processed through the inflaton decay to $N_iN_i$
which violates the lepton number.
If we remind that this $\sigma$ decay completes at 
$H\simeq \Gamma_\sigma$ for 
$\Gamma_\sigma=\sum_i\frac{\zeta_i^2}{8\pi}\tilde m_\sigma$, 
the reheating temperature could be estimated as\footnote{In the following 
part, the reheating temperature is fixed at (a) $T_R=5\tilde m_\sigma$ 
and (b) $T_R=2.5\tilde m_\sigma$.} 
\begin{equation}
T_R\simeq 1.74 g_\ast^{-1/4}\sqrt{M_{\rm pl}\Gamma_\sigma},
\label{rtemp}
\end{equation}
where $g_\ast=116$ is the relativistic degrees of freedom in the model.
Using this $T_R$, eq.~(\ref{cond2}) which is the condition for 
the symmetry restoration to be kept until the reheating time 
can be rewritten as
\begin{equation}
A(t_f)> 0.5 \frac{1}{g_S}
\left(\frac{u}{p_\ast}\right)^2 
\left(\frac{\tilde m_\sigma}{T_R}\right)^4.
\label{cond3}
\end{equation}  
Only the dilution of the number 
density due to the expansion of the Universe is taken into account 
in this condition.  
However, we should note that the breakdown of this symmetry restoration 
could be caused also by the decrease of the number density $n_{\pm S}$ 
due to the decay of $S_\pm$ to $N_iN_i$ which is caused by 
the interactions in eq.~(\ref{fmodel}).
This effect can be neglected as long as such a process decouples and
$\Gamma_{S_\pm}\ll H(T)$ is satisfied, where 
$\Gamma_{S_\pm}=\frac{\sum_iy_i^2}{8\pi}\tilde m_{S_\pm}$ and
$H(T)^2=\frac{\pi^2}{90}g_\ast\frac{T^{4}}{M_{\rm pl}^2}$.

In the left and right panels of Fig.~2, $\tilde m_S^2$ and 
$\frac{\Gamma_{S_\pm}}{H}$ (which is labeled by $D_S$) are 
plotted as the function of 
$x\left(\equiv\frac{M_1}{T}\right)$ at $T<T_R$, respectively.
These figures show that the restoration of the lepton number 
can be kept until a certain temperature $T^\prime$, which is lower than $T_R$.
The sudden decrease of $\frac{\Gamma_{S_\pm}}{H}$ is caused by the
threshold effects due to the generation of the Majorana mass of $N_i$
at $T^\prime$.  
In the right panel, we also plot the reaction rates of the $N_i$ 
scatterings mediated by the exchange of $S_\pm$ and $\sigma$ and 
also the inverse decay of $S_\pm$ and $\sigma$, which could wash out
the lepton number asymmetry since they violate the lepton number 
explicitly. 
These results show that any possible lepton number violating processes 
decouple at $T^\prime \le T\le T_R$.
If the lepton number asymmetry exists in the $N_i$ sector, 
it could be conserved at this stage since these processes are freezed out. 
This means that if the inflaton decay through the coupling 
$\sigma\bar N_i^cN_i$ could generate the lepton number asymmetry 
in the $N_i$ sector, it could be accumulated in the lepton sector 
where the lepton number is well defined and it is kept there until $T^\prime$.
  
A crucial problem is how the $CP$ symmetry could be violated in the 
inflaton decay. If its violation is realized at a substantial level, 
the lepton asymmetry generated through this decay 
could be distributed in the ordinary lepton sector through the lepton 
number conserving processes before reaching the symmetry breaking 
temperature $T^\prime$.
The sphaleron interaction can generate the baryon number asymmetry 
using this lepton number asymmetry.
In that case, the lepton number violating processes caused 
by the neutrino Yukawa coupling have to be sufficiently suppressed 
at $T<T^\prime$. It is crucial to avoid the washout of the non-thermally 
generated lepton number asymmetry in this scenario. 
If the initial lepton number asymmetry could take a sufficient value 
and satisfy these conditions, the scenario could be an alternative one 
to the thermal leptogenesis. 
It is worth studying whether this could give a
new possible scenario for non-thermal leptogenesis in
viable neutrino mass models.
In the next section, we take a radiative neutrino mass model as an example 
and propose a realistic framework for this leptogenesis scenario.

\section{Application to Leptogenesis}
\subsection{A particle physics model}
We consider an application of the symmetry restoration discussed in the
previous section to non-thermal leptogenesis in a one-loop radiative 
neutrino mass model, which is obtained by extending the Ma model \cite{ma} 
with singlet scalars.\footnote{Similar extension 
is studied in \cite{pseudod} in another context.}
It inherits favorable nature of the original Ma model, that is,
it can closely relate the neutrino mass generation to the dark matter (DM)
existence \cite{radnm}. 
The model is composed of an extra doublet scalar $\eta$, 
singlet fermions $N_i$, a real singlet scalar $\sigma$, and 
a complex singlet scalar $S$ in addition to the SM contents. 
We impose a $Z_2$ symmetry on the model and assign its odd parity both
$\eta$ and $N_i$. 
All other fields are assigned even parity including the 
inflaton $\sigma$ and $S$.
The Lagrangian relevant to these new contents contains 
the following terms,
\begin{eqnarray}
-{\cal L}&=&V_{\rm inf}(\sigma)
+g_\phi\sigma^2(\phi^\dagger\phi)
+g_\eta\sigma^2(\eta^\dagger\eta)+g_S\sigma^2(S^\dagger S)
+ \zeta_i\sigma\bar NN^c + \zeta_i^\ast\sigma\bar N^cN \nonumber \\
&+&\lambda_S\left(S^\dagger S -\frac{u^2}{2}\right)^2
+\frac{1}{2}m_S^2S^2+\frac{1}{2}m_S^2S^{\dagger 2}
+\kappa_\phi(S^\dagger S)(\phi^\dagger\phi)
+\kappa_\eta(S^\dagger S)(\eta^\dagger\eta) \nonumber \\
&+&y_iS\bar N_i^cN_i + y_i^{\ast}S^\dagger\bar N_iN_i^c 
+ h_{\alpha i}\bar\ell_\alpha N_i\eta   
+h_{\alpha i}^\ast\bar N_i\ell_\alpha\eta^\dagger \nonumber \\
&+&m_\phi^2\phi^\dagger\phi+m_\eta^2\eta^\dagger\eta 
+\lambda_1(\phi^\dagger\phi)^2 
+\lambda_2(\eta^\dagger\eta)^2 
+\lambda_3(\phi^\dagger\phi)(\eta^\dagger\eta)  
+\lambda_4(\eta^\dagger\phi)(\phi^\dagger\eta)  \nonumber \\
&+&\frac{\lambda_5}{2}\left[(\eta^\dagger\phi)^2  
+(\phi^\dagger\eta)^2 \right] ,
\label{smodel}
\end{eqnarray}
where $\ell_\alpha$ is the doublet lepton and $\phi$ is the ordinary 
doublet Higgs scalar.
A concrete form of $V_{\rm inf}(\sigma)$ is presented in the 
previous section.\footnote{If we apply Model (b) to this Lagrangian, 
$\sigma$ is just replaced by the complex $\sigma$ and $\sigma_{1,2}$ 
should be used in the study of the oscillation phenomena.}

This Lagrangian includes the potential (\ref{spot}) 
as a part of it.
However, the minimum of the potential for $S$ 
is shifted from $u^2$ to $\bar u^2\equiv u^2-\frac{m_S^2}{\lambda_S}$ 
because of the introduction of a new mass term $m_S^2S^2$. 
The masses of each component of $S$ and $\eta$ 
can be expressed as
\begin{equation}
m_{S_+}^2\simeq 2\lambda_S \bar u^2, \quad m_{S_-}^2\simeq -2m_S^2, \quad
M_{\eta^\pm}^2=\bar m_\eta^2+\lambda_3\langle\phi\rangle^2, \quad
M_{\eta_{R,I}}^2=\bar m_\eta^2+\lambda_\pm\langle\phi\rangle^2,
\label{mscalar}
\end{equation}
where $\bar m_\eta^2=m_\eta^2+\frac{\kappa_\eta}{2}\bar u^2$ and 
$\lambda_\pm=\lambda_3+\lambda_4\pm \lambda_5$.
The vacuum stability requires $m_S^2<0$ and
\begin{eqnarray}
&&\lambda_1,~\lambda_2,~\lambda_S>0, \quad
\lambda_3, \lambda_\pm>-\sqrt{\lambda_1\lambda_2}, \quad
\kappa_\phi>-\sqrt{\lambda_1\lambda_S}, \quad
\kappa_\eta>-\sqrt{\lambda_2\lambda_S}.
\label{stability}
\end{eqnarray}
The weak scale is derived as 
$\langle\phi\rangle^2=-\frac{1}{2\lambda_1}
\left(m_\phi^2+\frac{\kappa_\phi}{2}\bar u^2\right)$.
Since the Higgs mass is given as $m_h^2=4\lambda_1\langle\phi\rangle^2$, 
it imposes $\lambda_1\simeq 0.13$.
On the other hand, $\eta$ is assumed to have no vacuum expectation 
value (VEV) and then the $Z_2$ symmetry remains exact. 
As its result, neutrinos cannot get masses at tree level and 
the lightest $Z_2$ odd particle is stable.
This stable particle should be neutral to be a good DM candidate.
We take it as a neutral component ($\eta_R$) of $\eta$ here.
This imposes $\lambda_5<0$ and $\lambda_4+\lambda_5<0$. 

Before proceeding with further discussion, we order several comments 
relevant to the lepton number and its assignment to the new 
ingredients.\footnote{If we take another lepton number assignment, 
a different type of  
non-thermal leptogenesis could be considered \cite{ks2}.} 
Since the $\lambda_5$ term is indispensable for the small neutrino 
mass generation at the one-loop level as seen later \cite{pseudod,radnm}, 
$\eta$ should not have the lepton 
number as long as the lepton number conservation is imposed on this term.
As a result, $N_i$ should be assigned the 
lepton number 1 and then the coupling $S\bar N_i^cN_i$ requires 
that $S$ should have the lepton number $-2$ as discussed already. 
Unless the Majorana mass of $N_i$ is caused through 
the coupling $S\bar N_i^cN_i$ as a result of $\langle S\rangle\not=0$,
the neutrino mass cannot be generated at the low energy regions even at 
the loop level. Thus, 
the realization of $\langle S\rangle\not=0$ at low energy regions 
is required for the neutrino mass generation.
It should be also noted that the $Z_2$ symmetry is kept exact 
even after $S$ gets a VEV and then the existence of DM is guaranteed.
In the next part, we discuss neutrino masses and DM in this model. 

\subsection{Neutrino mass and dark matter}
Here we discuss the constraints derived from the low energy feature 
of the model after the breakdown of the symmetry restoration for $S$.
Neutrino masses are generated radiatively through one-loop diagrams 
with $N_i$ in the internal fermion line in the same way as the original 
Ma model. 
We apply the value of $\bar u$ in eq.~(\ref{vev}) to the right-handed 
neutrino masses $M_i= y_i\bar u$.
Since $M_{\eta_{R,I}}^2\gg |\lambda_5|\langle\phi\rangle^2$ is satisfied, 
the neutrino mass formula can be approximately written as
\begin{equation}
{\cal M}_{\alpha\beta}=\sum_i h_{\alpha i}h_{\beta i}\lambda_5\Lambda_i, \qquad
\Lambda_i\simeq \frac{\langle\phi\rangle^2}{8\pi^2M_i}
\ln\frac{M_i^2}{\bar M_\eta^2},
\label{lnmass}
\end{equation} 
where 
$\bar M_\eta^2= \bar m_\eta^2+(\lambda_3+\lambda_4)\langle\phi\rangle^2$. 
This suggests that the neutrino masses are obtained in almost the same way
as the ordinary seesaw model for $|\lambda_5|=O(1)$ in the present case.

In order to take account of the constraints from the neutrino 
oscillation data, we fix the flavor structure of
neutrino Yukawa couplings $h_{\alpha i}$ at the one which induces the 
tri-bimaximal mixing \cite{s},
\begin{equation}
h_{ej}=0, \quad h_{\mu j}=h_{\tau j}\equiv h_j \quad (j=1,2); \qquad 
h_{e3}=h_{\mu 3}=-h_{\tau 3}\equiv h_3. 
\end{equation}
In that case, the mass eigenvalues are estimated as 
\begin{eqnarray}
&&m_1=0, \qquad m_2= 3|h_3|^2\Lambda_3, \nonumber \\
&&m_3=2\left[|h_1|^4\Lambda_1^2+|h_2|^4\Lambda_2^2+
2|h_1|^2|h_2|^2\Lambda_1\Lambda_2\cos 2(\theta_1-\theta_2)
\right]^{1/2}, 
\end{eqnarray}
where $\theta_j={\rm arg}(h_j)$.
If we use $\bar u$ given in eq.~(\ref{vev}) and fix the parameters 
relevant to the neutrino masses as
\begin{eqnarray}
&&|h_1|= 0.1|\lambda_5|^{-1/2},   \qquad |h_2|= 0.38|\lambda_5|^{-1/2}, \qquad 
|h_3|= 0.15|\lambda_5|^{-1/2}, \nonumber \\
&&|y_1|= 0.1, \qquad |y_2|=0.12, \qquad 
|y_3|= 0.15, 
\label{yukawa}
\end{eqnarray}
the neutrino oscillation data could be explained.
Although a certain modification is required to reproduce the favorable 
mixing structure, it is sufficient for the study in the next section. 
We note that the smaller $|\lambda_5|$ requires the larger values of
neutrino Yukawa couplings.

The value of $|\lambda_5|$ is also constrained by the DM abundance.
In the present study, DM is assumed to be the real part $\eta_R$ 
of the neutral component of $\eta$. 
Its abundance could be tuned to the observed value 
as long as the couplings $\lambda_{3,4}$ take suitable values \cite{ks}.
Here, we should note that the allowed regions 
of $\lambda_3$ and $\lambda_4$ are 
constrained by eq.~(\ref{stability}) and the discussion below it.
Since $\bar m_\eta$ is assumed to be of $O(1)$~TeV, 
the mass of each component of $\eta$ is found to be degenerate enough
for the allowed values of $\lambda_{3,4}$ and $\lambda_5$. 
This makes the co-annihilation among them effective enough to
reduce the DM abundance sufficiently.
As an example, the expected relic abundance of $\eta_R$ for several 
values of $\lambda_{3,4}$ and $\bar m_\eta=1.75$~TeV is plotted in Fig.~3 
for the cases $\lambda_5=-1$ and $-0.5$.
The larger value of $\bar m_\eta$ is required for $|\lambda_5|~{^>_\sim}~1$.
In that case, the dependence of the relic abundance on $\lambda_{3,4}$
becomes much weaker compared to the case fixed by the smaller value of 
$|\lambda_5|$. The possible DM mass is strongly constrained 
to a narrow region depending on the value of $|\lambda_5|$. 
Anyway, the simultaneous explanation of the neutrino masses and 
the DM abundance could be preserved in this extended model. 
We should stress that no additional constraint from the neutrino physics 
and the DM physics is brought about by taking the present scenario. 
 
\begin{figure}[t]
\begin{center}
\epsfxsize=7.5cm
\leavevmode
\epsfbox{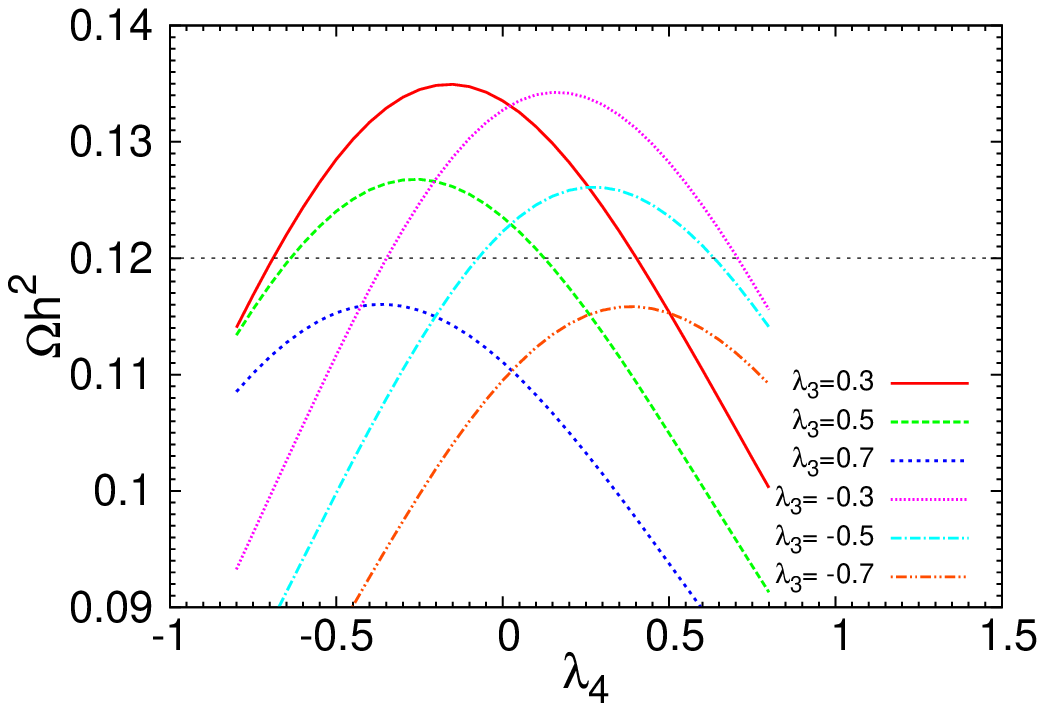}
\hspace*{5mm}
\epsfxsize=7.5cm
\leavevmode
\epsfbox{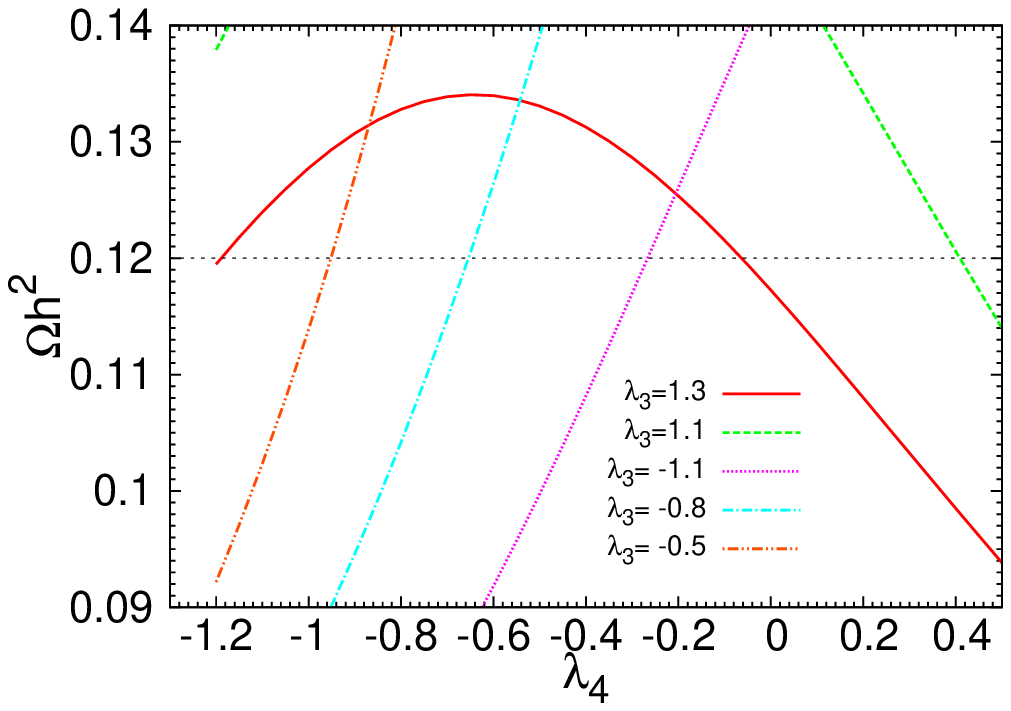}
\end{center}
%\vspace*{-3mm}

{\footnotesize {\bf Fig.~3}~~Relic abundance of $\eta_R$ in the case 
$\lambda_5=-1$ (left panel) and $-0.5$ (right panel). 
A horizontal dotted line $\Omega h^2=0.12$ is 
the required value from the observations \cite{planck15}.}
\end{figure}

It may be useful to give a remark for another aspect of the model. 
The VEV of $S$ could give the dominant origin for both the 
electroweak symmetry breaking and the DM mass through the interaction terms 
$\kappa_\phi S^\dagger S\phi^\dagger\phi$ and 
$\kappa_\eta S^\dagger S\eta^\dagger\eta$ unless they are 
forbidden by a certain reason.\footnote{If we assume the symmetry 
restoration due to 
the explosively produced $\eta$ or $\phi$, their couplings $\kappa_\eta$ 
or $\kappa_\phi$ with $S$ should take a substantial value as found 
from (\ref{effectivemass}). In that case, 
since they could induce large mass terms 
for $\eta$ and $\phi$ at the low energy region via the VEV $\bar u$, 
we could not adopt such a possibility in this model.
Only the explosive production of $S$ could not cause such a problem.} 
Since both scales of the electroweak symmetry breaking 
and the DM mass could be induced as $\kappa_\phi \bar u^2$ and 
$\kappa_\eta \bar u^2$ from the VEV $\bar u$, the couplings $\kappa_\phi$ and 
$\kappa_\eta$ should take negative and positive tiny values, respectively.
Such $\kappa_\phi$ and $\kappa_\eta$ satisfy 
the constraints given in eq.~(\ref{stability}).  
Although these couplings should take extremely small values for such a large 
value of $\bar u$ assumed in eq.~(\ref{vev}), 
it might present a possibility to 
unify the origin of the mass scales at TeV regions. 
These tiny couplings might be realized as non-renormalizable terms 
which are suppressed by the Planck mass, for example.

\subsection{Lepton asymmetry induced through the inflaton decay}
We consider the generation of the lepton number 
asymmetry through the decay of the inflaton $\sigma$ to a $N_i$ pair,
where the lepton number is supposed to be well defined.
This situation is also assumed to be kept until the generated asymmetry 
has been transferred from them to other particles. 
These assumptions require that $\langle S\rangle=0$ is satisfied 
throughout the period before the completion of the reheating at least.
In this conservative situation, the following study is done and then 
we need not take into account the washout of the generated asymmetry 
there.\footnote{We note that the washout processes caused through the
coupling $\sigma N_iN_i$ and $m_S^2S^2$ which break the lepton number 
explicitly are ineffective as shown in the right panel of Fig.~2.}
Such a situation cannot be realized in the case where the restoration of 
the lepton number is caused through the finite temperature effect. 
On the other hand, the symmetry restoration due to preheating discussed 
in the previous part can realize it as seen before. 

Here, it may be useful to compare the present scenario to the one 
discussed in \cite{b-l} previously in order to clarify the feature 
of the scenario.
In the latter model, the inflaton decays to the right-handed neutrinos 
nonthermally in which $U(1)_{B-L}$ is violated. The decay of these 
right-handed neutrinos generates the lepton number asymmetry. 
On the other hand, in the present model, the lepton number is considered 
to be generated through the inflaton decay to the right-handed neutrinos
where the lepton number $U(1)_L$ is assumed to be conserved and it is 
assumed to be kept until they decay to the doublet leptons.   

In order to generate the lepton number asymmetry through 
the lepton number violating decay of the inflaton $\sigma$ to $N_iN_i$, 
the $CP$ violation is required there.
The mass term $m_S^2 S^2$ in the second line 
of eq.~(\ref{smodel}), which breaks the lepton number 
explicitly,\footnote{This explicit breaking of the lepton 
number makes a Nambu-Goldstone boson caused by its spontaneous symmetry
breaking (SSB) heavy enough as shown in eq.~(\ref{mscalar}). 
The topological defect which could appear through this SSB 
is not stable due to the same explicit breaking.} 
can play a crucial role for this $CP$ violation.
On the other hand, since the lepton number violation in the $S$ sector 
also causes the washout of the generated lepton number asymmetry through the 
scattering, it has to be taken into account in 
the estimation of the final lepton number asymmetry.
Related to this point, we should remember that the symmetry 
restoration due to the preheating could be much more effective compared 
to the one due to the finite temperature effect of the reheating 
\cite{sym-rest1}. As its result, these violating effects could be freezed 
out throughout the symmetry restored period as seen in the right panel 
of Fig.~2.  

\begin{figure}[t]
\begin{center}
\epsfxsize=14cm
\leavevmode
\epsfbox{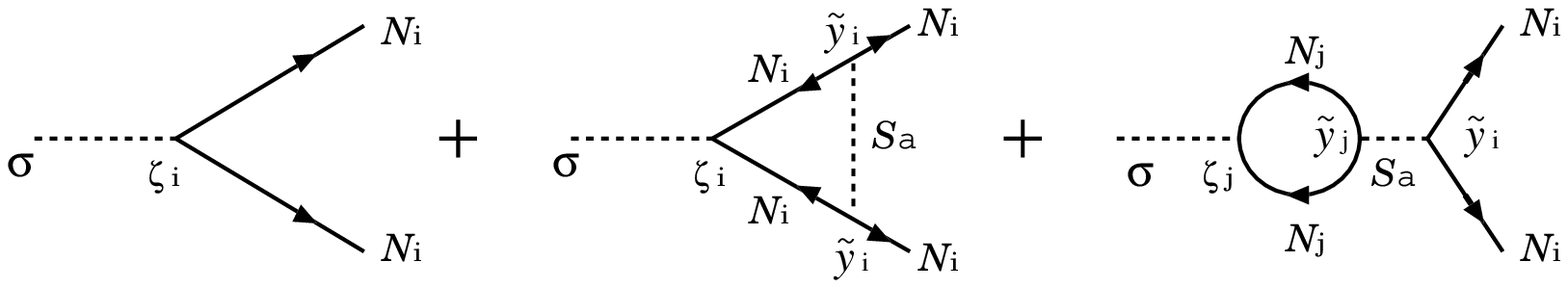}
%\vspace*{-3mm}
\end{center}

{\footnotesize {\bf Fig.~4}~~Feynman diagrams contributing to 
the generation of lepton number asymmetry. $S_a$ stands for 
the mass eigenstates $S_\pm$ and the couplings $\tilde y_i$ are
fixed at the ones shown in eq.(\ref{scoup}).}
\end{figure}

Since the lepton number is violated in the interaction which causes the 
inflaton decay, the lepton number asymmetry could be generated
if $CP$ is violated in this process.
In order to see how the $CP$ could be violated there, 
we note that $S$ is decomposed into two mass 
eigenstates $S_\pm$ by the explicit lepton number violation due 
to the mass term $m_S^2S^2$ even at the symmetry restored period. 
Their mass eigenvalues are $m_{S_\pm}^2=\tilde m_S^2 \pm m_S^2$,
where $\tilde m_S^2$ is the mass brought about through the symmetry 
restoration due to the preheating. It is given by
$\tilde m_S^2\simeq \lambda_S\left(\frac{2n_S(t)}{p_m}-\bar u^2\right)$
as found from eq.~(\ref{effectivemass}).
If we use these mass eigenstates,
the couplings of $N_i$ and $S$ in eq.~(\ref{smodel}) can be rewritten as
\begin{equation}
y_iS\bar N_i^cN_i + y_i^{\ast}S^\dagger\bar N_iN_i^c 
=\frac{1}{\sqrt 2}y_iS_+\bar N_i^cN_i+\frac{i}{\sqrt 2}y_iS_-\bar N_i^cN_i
+\frac{1}{\sqrt 2}y_i^\ast S_+\bar N_iN_i^c 
-\frac{i}{\sqrt 2}y^\ast_i S_-\bar N_iN_i^c.
\label{scoup}
\end{equation}
The $CP$ violation in the inflaton decay
could be caused from the interference between the tree diagram and 
the one-loop diagram which is induced by these couplings as shown in Fig.~4.

The $CP$ asymmetry $\varepsilon$ in this inflaton decay is defined as
\begin{equation}
\varepsilon=\frac{\Gamma(\sigma\rightarrow \sum_iN_iN_i)-
\Gamma(\sigma\rightarrow \sum_iN_i^cN_i^c)}
{\Gamma(\sigma\rightarrow \sum_iN_iN_i)+
\Gamma(\sigma\rightarrow \sum_iN_i^cN_i^c)}.
\end{equation}
Since the contribution from the self-energy diagram in Fig.~4 is 
negligible for non-degenerate values of $\tilde m_\sigma^2$ and $\tilde m_S^2$, 
we find that $\varepsilon$ could be expressed as
\begin{eqnarray}
(a)&& \varepsilon=\frac{1}{4\pi}
\frac{m_S^2}{\tilde m_\sigma^2}
\frac{ \sum_i{\rm Im}(\zeta_i^2y_i^{\ast 2})}
{\sum_i|\zeta_i|^2}\sim \frac{\sum_i|y_i|^2}{12\pi}
\frac{m_S^2}{\tilde m_\sigma^2}, \nonumber \\
(b)&& \varepsilon=\frac{1}{2\pi}
\frac{m_\sigma^2m_S^2}{m_{+\sigma}^2m_{-\sigma}^2}
\frac{\sum_i{\rm Im}(\zeta_i^2y_i^{\ast 2})}
{\sum_i|\zeta_i|^2}
\sim \frac{\sum_i|y_i|^2}{6\pi}
\frac{m_\sigma^2m_S^2}{\tilde m_\sigma^4}, 
\label{cp}  
\end{eqnarray}
where the maximal $CP$ phase and the universality of $\zeta_i$ 
are assumed in the last expressions for each model.  
These formulas show that $\varepsilon$ is proportional to the
mass difference between $S_+$ and $S_-$ in both models. 
It is also proportional to the mass difference 
between $\sigma_1$ and $\sigma_2$ in Model (b). 
Thus, these mass differences $m_S^2$ and
$m_\sigma^2$ should not be so small compared to $\tilde m_\sigma^2$ 
in order to guarantee a sufficient value for the $CP$ asymmetry 
$\varepsilon$.

\begin{figure}[t]
\begin{center}
\begin{tabular}{ccccccc}\hline
  & $x_R~\left(\equiv\frac{M_1}{T_R}\right)$ & 
$x^\prime~\left(\equiv\frac{M_1}{T^\prime}\right)$& $\tilde m_S(x_R)$
& $\varepsilon$ & $Y_L(x_R)$ \\ \hline
Model (a)& $6.7$& $10.3$ &$3.4\times 10^{10}$  
& $1.4\times 10^{-8}$ & $6.5\times 10^{-8}$   \\ 
Model (b)& $10$ & $13.8$ & $4.0\times 10^{10}$ 
& $1.4\times 10^{-9}$ & $3.3\times 10^{-9}$  
\\ \hline
\end{tabular}
\end{center}
\vspace*{5mm}

{\footnotesize {\bf Table~2}~~Results obtained through the numerical study 
in each model defined by the parameters in Table~1. 
The value of $x^\prime$ can be read from Fig.~2. In both models, 
$M_1$ is fixed at $10^{14}$~GeV.  }
\end{figure}

Taking account of the arguments presented by now, 
we can summarize the necessary conditions for the lepton number asymmetry
generated in this scenario to be the origin of the baryon number 
asymmetry in the Universe as follows:\\
(i) The symmetry restoration should break down after the
completion of reheating. This requires that $S$ gets the VEV 
at $T^\prime$ which is smaller than $T_R$.
If it is not satisfied, the asymmetry generated before the symmetry 
breaking is erased by the thermalization at the reheating.\\   
(ii) The Majorana mass $M_i=y_i\bar u$ generated through the symmetry breaking 
should satisfy $M_i> T^\prime$ or the neutrino Yukawa couplings have to
be small enough.\footnote{Since these conditions should be satisfied 
consistently with the explanation of neutrino oscillation data, 
the study in the previous part shows the latter one is not allowed 
in the present model.}
Otherwise, since the lepton number violating processes containing $N_i$ 
could be in thermal equilibrium, the existing lepton number asymmetry 
is washed out immediately through these processes \cite{s}.
In that case, the initial lepton number asymmetry plays no role and 
the scenario is reduced to the usual thermal leptogenesis.\\
(iii) The inflaton mass and the effective mass of $S$ caused by the 
symmetry restoration due to the preheating should satisfy 
$\tilde m_\sigma,~\tilde m_S \gg T^\prime$.
Since the $N_i$ scatterings mediated by the exchange of $\sigma$ and $S_\pm$
violate the lepton number, they have to be freezed out to keep the 
asymmetry generated in the $N_i$ sector. \\
If these conditions are satisfied, the lepton number asymmetry generated
in the $N_i$ sector is expected to be immediately distributed to 
the SM contents by the interactions which could be 
in the thermal equilibrium at $T_R$.

We introduce the lepton asymmetry 
in the comoving volume as $Y_L\equiv \frac{n_L}{s_R}$
by using the entropy density $s_R=\frac{2\pi^2}{45}g_\ast T_R^3$,
where $n_L$ is defined as the difference between the lepton number 
density and the antilepton number density.
It could be estimated at $T_R$ as
\begin{equation}
Y_L(T_R)=\frac{2\varepsilon n_\sigma(T_R)}{s_R}\simeq 1.5\varepsilon
\frac{T_R}{\tilde m_\sigma},
\label{init}
\end{equation} 
where $n_\sigma$ is defined as $n_\sigma=\frac{\rho_\sigma}{\tilde m_\sigma}$
by using $\rho_\sigma$ which is the energy density of $\sigma$ and  
determined by $H\simeq \Gamma_\sigma$.  
The baryon number asymmetry is generated through the conversion 
of this lepton number asymmetry $Y_L$ by the $B-L$ conserving 
sphaleron interaction. 
If we solve the equilibrium conditions for the chemical potential, 
the baryon number asymmetry is found to be obtained as 
$Y_B=-\frac{8}{15}Y_L$ in this model.
Thus, the present $Y_B$ is calculated from $Y_L(T_{\rm EW})$ where $T_{\rm EW}$ 
is the sphaleron decoupling temperature $T_{\rm EW}\simeq 100$~GeV.
The evolution of the lepton number asymmetry after the
breaking of the symmetry restoration at $T^\prime$ follows the Boltzmann 
equations. In that study, 
we can use the lepton number asymmetry $\frac{5}{8}Y_L(T_R)$ in the ordinary 
doublet leptons as an initial value for $Y_L$ at $T^\prime$. 
It could take a sufficient value 
only if the scalar mass differences are not strongly suppressed.
For example, they should satisfy $m_S^2>10^{-7}\tilde m_\sigma^2$ in 
Model (a) and $m_S^2m_\sigma^2>10^{-7}\tilde m_\sigma^4$ in Model (b) 
for $T_R\sim \tilde m_\sigma$ and $|y_i|\simeq 0.1$. 

\begin{figure}[t]
\begin{center}
\epsfxsize=7.5cm
\leavevmode
\epsfbox{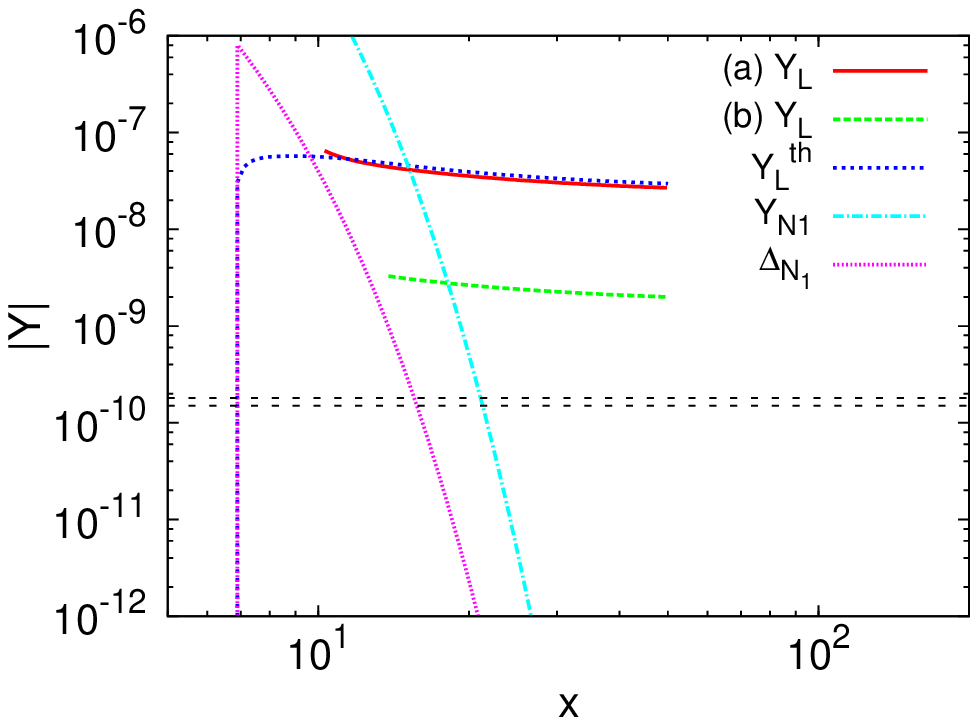}
\hspace{5mm}
\epsfxsize=7.5cm
\leavevmode
\epsfbox{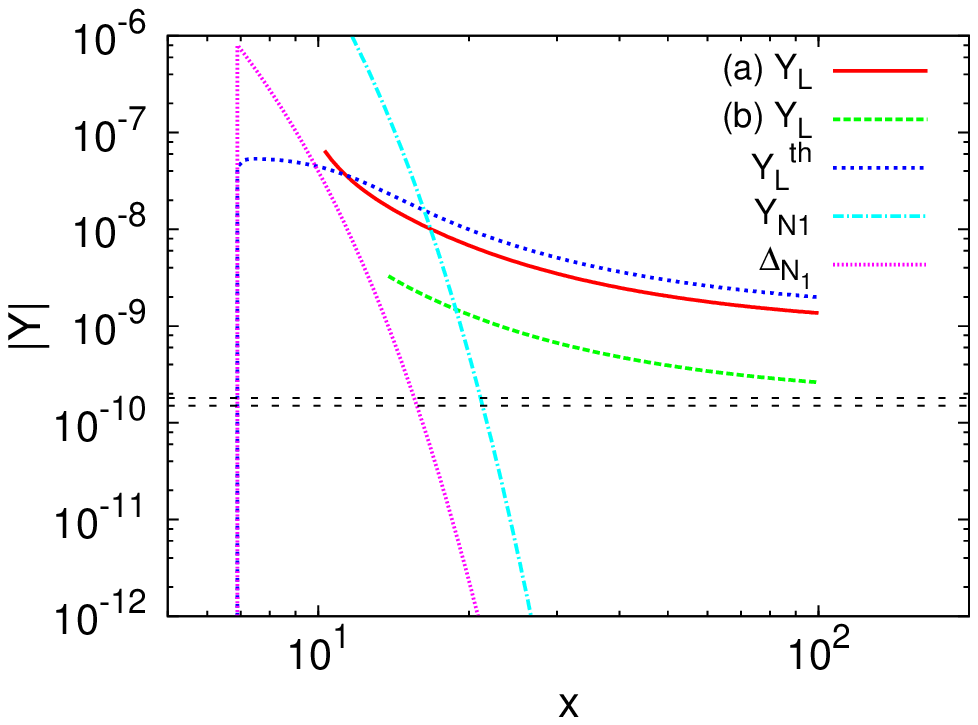}
\end{center}
%\vspace*{-3mm}
{\footnotesize {\bf Fig.~5}~~The evolution of the lepton number 
asymmetry $Y_L$ at $x>x^\prime$, which is obtained as the solution of 
the Boltzmann equations for Models (a) and (b).
As a reference, $Y_L^{\rm th}$, $Y_{N_1}$ and 
$\Delta_{N_1}(\equiv \left|Y_{N_1}-Y_{N_1}^{\rm eq}\right|)$ for 
the thermal leptogenesis in Model (a) are plotted at $x>x_R$.   
The value of $Y_L$ required to explain the observed baryon number 
asymmetry is shown as the range sandwiched by 
the horizontal dotted lines. The left and right panels correspond to
$\lambda_5=-1$ and $-0.5$, respectively.}
\end{figure}

In order to estimate $Y_L(T_{\rm EW})$ correctly, it is necessary to take into 
account the washout effect of the lepton number asymmetry at $T<T^\prime$.
It is induced through the inverse decay and the scattering processes which 
include $N_i$ in them. 
One may consider a situation such 
as $M_i\gg T_R$ as a specific situation. 
Since the washout effects could be almost freezed out at $T_R$ in this case, 
we can expect $Y_L(T_{\rm EW})\simeq Y_L(T_R)$.
Thus, the baryon number asymmetry is determined as
$|Y_B|\simeq 0.5 \varepsilon \frac{T_R}{\tilde m_\sigma}$.
On the other hand, in the marginal case $M_i~{^>_\sim}~T^\prime$, 
the washout effects are crucial and we need to solve the Boltzmann 
equations which include their effects appropriately. 
The relevant Boltzmann equations at $T\le T^\prime$ are given as \cite{boltz}
\begin{eqnarray}
&&\frac{dY_{N_1}}{dx}=-\frac{x}{sH(M_1)}
\left(\frac{Y_{N_1}}{Y_{N_1}^{\rm eq}}-1\right)\left\{
\gamma_D^{N_1}+\sum_{j=1,2}\left(\gamma_{N_1N_j}^{(2)}+
\gamma_{N_1N_i}^{(3)}\right)\right\}, \nonumber \\
&&\frac{dY_L}{dx}=\frac{x}{sH(M_1)}\left\{
\varepsilon_{N_1}\left(\frac{Y_{N_1}}{Y_{N_1}^{\rm eq}}-1\right)
\gamma_D^{N_1}
-\frac{2Y_L}{Y_\ell^{\rm eq}}\left(\frac{\gamma_{ID}^{N_{2,3}}}{4}+
\gamma_N^{(2)} +\gamma_N^{(13)}\right)\right\}, 
\label{bqn}
\end{eqnarray}
where a hierarchical right-handed neutrino mass spectrum is assumed.
$x$ is a dimensionless variable defined as $x=\frac{M_1}{T}$ and 
$N_1$ stands for the lightest one. 
In these equations, we include the decay of $N_1$ ($\gamma_D^{N_1}$), 
the inverse decay of $N_{2,3}$ ($\gamma_{ID}^{N_{2,3}}$), 
and the lepton number violating scatterings mediated by $\eta$ 
($\gamma_N^{(2)}$) and by $\ell_\alpha$ ($\gamma_N^{(13)}$). 
The expression of each reaction density $\gamma$ can be found in \cite{ks}.

The initial values of $Y_L$ for these Boltzmann equations are given 
in Table~2, which are obtained for the parameters used in the symmetry 
restoration study in the previous part. 
The results of the numerical calculation are shown in Fig.~5 
in the cases $\lambda_5=-1$ (left panel) and $\lambda_5=-0.5$ 
(right panel).
In this study, the $CP$ asymmetry in the $N_1$ decay is assumed to be
$\varepsilon_{N_1}=-4.0\times 10^{-8}$, although it can take
$|\varepsilon_{N_1}|=O(10^{-3})$ for the maximal $CP$ phase.
This allows us to neglect the lepton number asymmetry generated 
by the thermal origin in the final result. 
Since $x^\prime>10$ is satisfied in both cases, 
the Boltzmann suppression is effective for the lepton number 
violating processes.
On the other hand, the neutrino oscillation data require
that the neutrino Yukawa couplings should not be so small and 
of $O(10^{-1})$ as found from eq.~(\ref{yukawa}) 
since the right-handed neutrinos are heavy.
As a result, the decoupling of the lepton number violating processes 
could be marginal.  
The figure shows that $Y_L(x_{\rm EW})\simeq Y_L(x^\prime)$ is satisfied
for $\lambda_5=-1$ in both models.
In the case $\lambda_5=-0.5$, however, their decoupling is not sufficient
and then $Y_L$ decreases gradually to a fixed value.
Since the lepton number violating processes
sufficiently decouple at $x\gg x^\prime$, the lepton number asymmetry 
$Y_L$ could keep a substantial value until $x_{\rm EW}$.
In the same panels, as a reference, we also plot the results of the 
thermal leptogenesis for the same parameter sets but the 
initial values such as $Y_L(x^\prime)=0$ and 
$Y_{N_1}(x^\prime)=Y_{N_1}^{\rm eq}(x^\prime)$.
The lepton number asymmetry produced through it is found to 
take the same order values as the non-thermal case.
This is because $|\varepsilon_{N_1}|\Delta_{N_1}> 10^{-10}$ 
is satisfied at $x~{^>_\sim}~x^\prime$ for $|\varepsilon_{N_1}|=O(10^{-3})$,
which is realized for the maximal $CP$ phase.
If the $CP$ phase in the neutrino Yukawa couplings does not 
take such a large value, $\varepsilon_{N_1}$ could not be large enough
and the thermal leptogenesis could not produce the required 
baryon number asymmetry.
This condition is not required for the present non-thermal scenario.
It is irrelevant to the $CP$ phase in the neutrino Yukawa couplings.
Thus, the present non-thermal leptogenesis 
scenario could be an alternative origin for the baryon number 
asymmetry in the Universe under such a situation. 
We should note that the scenario is closely related to the neutrino mass 
generation and the DM candidate in a somewhat different way 
from the thermal leptogenesis.

\section{Summary}  
We have proposed a scenario for the non-thermal leptogenesis 
associated to the reheating due to the inflaton decay.
If inflaton is assumed to couple with the right-handed neutrinos, 
its out-of-equilibrium decay might generate
the lepton number asymmetry in the right-handed neutrinos as long as
the lepton number is conserved in this sector at such a period.
The lepton number asymmetry generated in the right-handed neutrino sector 
could be transferred to the doublet lepton sector through the 
lepton number conserving decay.
If the transferred asymmetry could take a substantial value, 
the sufficient baryon number asymmetry is expected to be generated 
from it. 
On the other hand, at low energy regions the lepton number violation 
in the right-handed neutrino sector is necessary for the neutrino 
mass generation.
Thus, the lepton number should be restored at the era of the inflaton 
decay for this scenario to work well.
Preheating associated to the inflation might realize such symmetry 
restoration. 

In this paper, we have studied such a possibility and its application 
to a one-loop radiative neutrino mass model extended 
by the singlet scalars. If the inflaton is a singlet scalar,
it could have the couplings necessary for this scenario in general.
The present study shows that the model can explain the neutrino masses,
the DM abundance and the baryon number asymmetry in the Universe 
simultaneously.
The scenario might be applicable for other various particle models. 
Especially, the ordinary seesaw model could be such a candidate since
the right-handed neutrino masses are in a favorable range in the 
inflation models studied here. 
However, the DM cannot be included in that case.
In this direction, it may be an interesting subject to combine it 
with an axion DM model.

\section*{Acknowledgements}
This work is partially supported by MEXT Grant-in-Aid 
for Scientific Research on Innovative Areas (Grant No. 26104009).

\newpage
\bibliographystyle{unsrt}

\begin{thebibliography}{99}
\bibitem{uobs}WMAP Collaboration, D.~N.~Spergel, {\it et al.},
Astrophys. J. Suppl.  {\bf 148} (2003) 175; 
E.~Komatsu, {\it et al.}, Astrophys. J. Suppl. {\bf 180} (2009) 330; 
E.~Komatsu, {\it et al.}, Astrophys. J. Suppl. {\bf 192} (2011) 18;
G.~Hinshaw, {\it et al.}, Astrophys. J. Suppl. {\bf 208} (2013) 19;
Planck Collaboration, P.~A.~R.~Ade, {\it et al.}, Astron. Astrophys. 
{\bf 571} (2014) A22.

\bibitem{planck15}Planck Collaboration, P.~A.~R.~Ade, {\it et al.},
Astron. Astrophys. {\bf 594} (2016) A13; 
Planck Collaboration, P.~A.~R.~Ade, {\it et al.}, 
Astron. Astrophys. {\bf 594} (2016) A20.

\bibitem{preheat1}L.~Kofman, A.~Linde and A.~A.~Starobinsky,
 Phys. Rev. Lett. {\bf 73} (1994) 3195; Phys. Rev. {\bf D56} (1997) 3258.

\bibitem{sym-rest1}L.~Kofman, A.~Linde and A.~A.~Starobinsky, 
Phys. Rev. Lett. {\bf 76} (1996) 1011; 
I.~I.~Tkachev, Phys. Lett. {\bf B376} (1996) 35. 

\bibitem{sym-rest2}A.~Riotto and I.~I.~Tkachev, Phys. Lett. 
{\bf B385} (1996) 57. 

\bibitem{pre-baryog}G.~W.~Anderson, A.~Linde and A.~Riotto, Phys. Rev. 
Lett. {\bf 77} (1996) 3716,
E.~W.~Kolb, A.~Linde and A.~Riotto, Phys. Rev. Lett. 
{\bf 77} (1996) 4290.

\bibitem{pre-phase}S.~Khlebnikov, L.~Kofman, A.~Linde and I.~Tkachev, 
Phys. Rev. Lett. {\bf 81} (1998) 2012,
A.~Rajantie and E.~J.~Copeland, Phys. Rev. Lett. {\bf 85} (2000) 916.

\bibitem{pre-sec-inf}G.~Felder, L.~Kofman, A.~Linde and I.~Tkachev, 
JHEP {\bf 08} (2000) 010,

\bibitem{pre-tach} G.~Felder, L.~Kofman and A.~Linde, Phys. Rev. {\bf D64} 
(2001) 123517.

\bibitem{hybrid}A.~D.~Linde, Phys. Rev. {\bf D49} (1994) 748.

\bibitem{therm}H.~Umezawa, H.~Matsumoto and M.~Tachiki, 
{\it Thermo Field Dynamics and Condensed States}, North Holland, 1982; 
P.~A.~Henning, Phys. Rep. {\bf 253} (1995) 235.

\bibitem{dj}L.~Dolan and R.~Jackiw, Phys. Rev. {\bf D9} (1974) 3320.

\bibitem{nonm-inf}B.~L.~Spokoiny, Phys. Lett. {\bf 147B} (1984) 39; 
D.~S.~Salopek, J.~R.~Bond and J.~M.~Bardeen, 
Phys. Rev. {\bf D40} (1989) 1753.    

\bibitem{higgsinf}F.~L.~Bezrukov and M.~Shaposhnikov, 
Phys. Lett. {\bf B659} (2008) 703; 
F.~L.~Bezrukov, A.~Magnin and M.~Shaposhnikov, Phys. Lett. 
{\bf B675} (2009) 88.

\bibitem{sinfl}R.~N.~Lerner and J.~McDonald, Phys. Rev. {\bf D80} 
(2009) 123507; R.~N.~Lerner and J.~McDonald, Phys. Rev. {\bf D83} 
(2011) 123522.

\bibitem{bks1}R.~H.~S.~Budhi, S.~Kashiwase and D.~Suematsu,
Phys. Rev. {\bf D93} (2016) 013022.

\bibitem{bks2}R.~H.~S.~Budhi, S.~Kashiwase and D.~Suematsu,
Phys. Rev. {\bf D90} (2014) 113013; JCAP {\bf 09} (2015) 039. 

\bibitem{mc}J.~McDonald, JCAP 09 (2014) 027.

\bibitem{ma}E.~Ma, Phys. Rev. {\bf D73} (2006) 077301.

\bibitem{pseudod}S.~Kashiwase and D.~Suematsu, Eur. Phys. 
J {\bf C76} (2016) 117. 

\bibitem{radnm}J.~Kubo, E.~Ma and D.~Suematsu, 
Phys. Lett. {\bf B642} (2006) 18; J.~Kubo and D.~Suematsu, 
Phys. Lett. {\bf B643} (2006) 336; D.~Suematsu, Eur. Phys. 
J. {\bf C56} (2008) 379; D.~Aristizabal Sierra, J.~Kubo, D.~Suematsu,
D.~Restrepo and O.~Zapata, Phys. Rev. {\bf D79} (2009) 013011;
D.~Suematsu, T.~Toma and T.~Yoshida, Phys. Rev. {\bf D79} (2009) 093004; 
D.~Suematsu, T.~Toma and T.~Yoshida, Phys. Rev. {\bf D82} (2010) 013012; 
D.~Suematsu, Eur. Phys. J {\bf C72} (2012) 1951. 

\bibitem{ks2}D.~Suematsu, Phys. Rev. {\bf D85} (2012) 073008;
S.~Kashiwase and D.~Suematsu, Phys. Lett. {\bf B749} (2015) 603.

\bibitem{s}D.~Suematsu, Phys. Lett. {\bf B760} (2016) 538.

\bibitem{ks}S.~Kashiwase and D.~Suematsu, Phys. Rev. {\bf D86} (2012)
	053001; S.~Kashiwase and D.~Suematsu, Eur Phys. J {\bf C73} (2013)
	2484. 

\bibitem{b-l}W.~Buchmuller, K.~Schmitz and G.~Vertongen, 
Phys. Lett. {\bf B693} (2010) 421.

\bibitem{boltz}M.~A.~Luty, Phys. Rev. {\bf D45} (1992)455;
M.~Pl\"umacher, Z. Phys. {\bf C74} (1997)549.
\end{thebibliography}

\end{document}